# BreastSAM: A Study of Segment Anything Model for Breast Tumor Detection in Ultrasound Images

Mingzhe Hu[a], Yuheng Li[b] and Xiaofeng Yang[a,b,c*]

[a]Department of Computer Science and Informatics, Emory University, GA, Atlanta, USA
[b]Department of Biomedical Engineering, Emory University, GA, Atlanta, USA
[c]Department of Radiation Oncology, Winship Cancer Institute, School of Medicine, Emory University, GA, Atlanta, USA

*Email: xiaofeng.yang@emory.edu

## Abstract

Breast cancer is one of the most common cancers among women worldwide, with early detection significantly increasing survival rates. Ultrasound imaging is a critical diagnostic tool that aids in early detection by providing real-time imaging of the breast tissue. We conducted a thorough investigation of the Segment Anything Model (SAM) for the task of interactive segmentation of breast tumors in ultrasound images. We explored three pre-trained model variants: ViT_h, ViT_l, and ViT_b, among which ViT_l demonstrated superior performance in terms of mean pixel accuracy, Dice score, and IoU score. The significance of prompt interaction in improving the model's segmentation performance was also highlighted, with substantial improvements in performance metrics when prompts were incorporated. The study further evaluated the model's differential performance in segmenting malignant and benign breast tumors, with the model showing exceptional proficiency in both categories, albeit with slightly better performance for benign tumors. Furthermore, we analyzed the impacts of various breast tumor characteristics - size, contrast, aspect ratio, and complexity - on segmentation performance. Our findings reveal that tumor contrast and size positively impact the segmentation result, while complex boundaries pose challenges. The study provides valuable insights for using SAM as a robust and effective algorithm for breast tumor segmentation in ultrasound images.

# 1. Introduction

Breast tumor, also known as a breast lump, refers to the abnormal growth of tissue within the breast. It is typically characterized by partial swelling and thickening of the breast tissue, as well as the formation of a lump in the breast or under the armpit[1]. Breast tumors can be broadly categorized into two types based on their nature: benign tumors or malignant tumors (breast cancer). Benign breast tumors usually have well-defined borders and exhibit a round or oval shape on medical imaging. On the other hand, malignant tumors appear irregular and may have lobules [2]. While benign breast tumors generally do not pose a life-threatening risk, they can increase the chances of developing breast cancer in the future. According to the latest statistics[3] from 2022, more than 250,000 women are diagnosed with breast cancer in the United States each year, with the majority of cases occurring in women aged 50 and above (accounting for over 90% of cases). Throughout their lifetime, approximately one in every eight women has a possibility of developing invasive breast cancer, which carries a mortality rate of approximately 2.5%. Timely prevention and treatment of breast cancer can greatly improve patients' survival rates. Currently, ultrasound imaging is the primary imaging technique for breast cancer. Accurately determining the boundaries and extent of breast cancer tissue is crucial for downstream tasks such as surgery. However, the current process heavily relies on manual extraction of the breast cancer region from the images, which poses certain limitations and challenges. The manual extraction method is time-consuming and requires significant expertise and experience from medical professionals. It is a subjective process that can be prone to inter-observer variability, where different doctors may interpret the images differently and extract the cancer region with varying degrees of accuracy. To overcome these issues, establishing a pipeline that assists doctors in automating or semi-automating the segmentation of breast cancer would be a significant advancement. Such a pipeline can enhance the accuracy and efficiency of diagnosis, providing better treatment options for patients.

Over the last ten years, the rapid advancement of frameworks like R-CNN[4], U-Net[5], mask scoring CNN[6], vision transformers[7-9], and mlp-mixer[10] has significantly transformed the field of medical image segmentation, demonstrating superior performance[11-14] in contrast to traditional segmentation techniques. Nonetheless, there are still hurdles to overcome with both supervised and unsupervised models built on these frameworks. Supervised learning techniques typically require a sufficient quantity of validated labels during the training phase. However, there's a scarcity of extensive, high-quality, annotated datasets for medical images, primarily due to the high levels of human labor and expert knowledge needed for the labeling process. Furthermore, the effectiveness of models trained on a large dataset from one institution often diminishes when extended to other institutions, raising doubts about their practical applicability. While there are unsupervised methods available, they lack definitive segmentation masks, making their accuracy and dependability less certain compared to their supervised counterparts.

The emergence of Meta's Segment Anything Model (SAM)[15], one of the inaugural attempts at a foundational model for computer vision tasks, has significantly shifted the landscape. The hallmark of a foundational model lies in its capacity to handle tasks and data it hasn't encountered before - a prime example of zero or few-shot learning. Drawing inspiration from language models[16], which enable zero or few-shot learning on new datasets and tasks through prompting, Meta designed the concept of promptable segmentation as the pre-training task. Here, the model is expected to generate a valid segmentation given a prompt, which could be a point, a bounding box, a mask, or even text. Even in the face of ambiguous prompts, a valid segmentation is required, pushing the model towards greater generalization.

This research endeavored to adapt the SAM originally designed for natural images to breast ultrasound images, with the aim of achieving precise breast tumor segmentation. While previous attempts[17-21] had been made to employ SAM for breast cancer segmentation, they overlooked the differentiation between benign and malignant tumors and did not explore the influence of morphological features on the segmentation performance of breast tumors. Our study made significant contributions in four key areas.

- Firstly, we conducted a thorough performance comparison of three distinct pretrained model structures: ViT-b, ViT-l, and ViT-H. By analyzing their respective capabilities, we ascertained the most effective approach for breast tumor segmentation.
- Secondly, we delved into the significance of prompt instructions for SAM, comprehending the essential role they played in guiding the segmentation process. Our investigation shed light on the importance of prompts in achieving accurate and reliable results.
- Additionally, we meticulously examined the disparities in performance between the segmentation of benign and malignant tumors. By unraveling these distinctions, we gained valuable insights into the challenges associated with distinguishing between these two tumor types.
- Lastly, our research critically explored the impact of key tumor characteristics on the segmentation results. By identifying and comprehending these essential features, we enhanced our understanding of the factors that influenced the accuracy and reliability of breast tumor segmentation.

By addressing these pivotal aspects, our research not only contributes to advancing the field of breast tumor segmentation in ultrasound images but also enhances our diagnostic capabilities and aids in treatment planning for breast cancer patients.

## 2. Methods

2.1 Breast Tumor Ultrasound Dataset

The dataset[22] used in this study consists of breast ultrasound images collected in 2018 from 600 female patients, ranging in age from 25 to 75 years. The dataset contains 780 PNG images, with an average size of 500 x 500 pixels. The images are categorized into three classes: normal, benign, and malignant. The distribution of images in each class is as follows: 133 normal, 487 benign, and 210 malignant. The dataset was collected and stored in DICOM format at Baheya Hospital, and after preprocessing, the number of images was reduced to 780. The images were obtained using LOGIQ E9 ultrasound systems and LOGIQ E9 Agile ultrasound systems, with a resolution of 1280 x 1024. Ground truth or mask images were created for each image using MATLAB. Figure 1 shows example images of this dataset. For segmentation tasks, we only used the benign and malignant images.

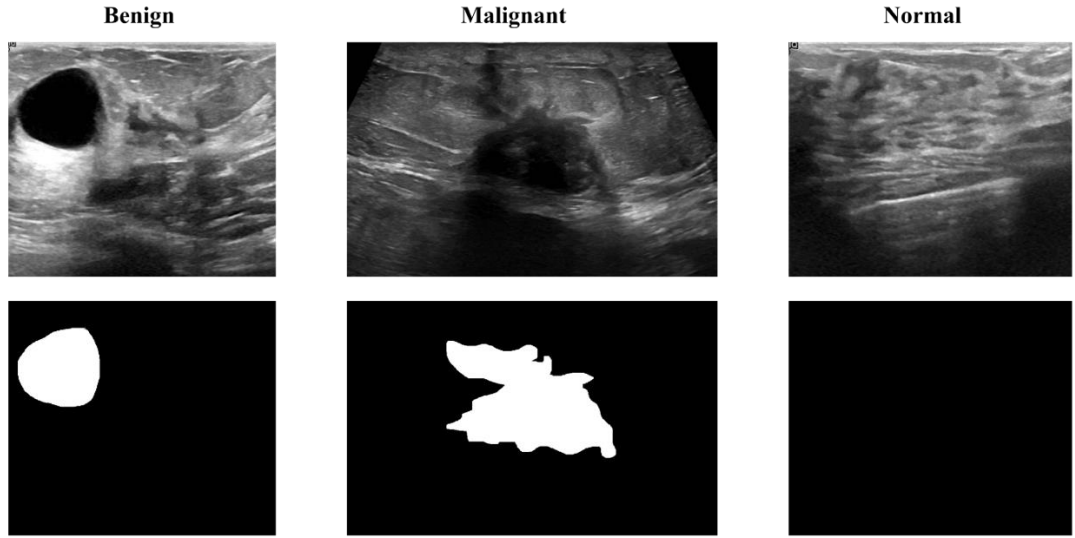

**Figure 1**: Example images from the dataset of breast ultrasound dataset and their corresponding masks. The top row represents the breast ultrasound images, while the bottom row displays the corresponding masks. The three columns represent the categories of benign, malignant, and normal cases.

2.2 Segment Anything Model (SAM)

SAM's Image Encoder utilizes a ViT pre-trained through Masked Auto Encoding (MAE), converting a 1024x1024x3 dimensional image into 64x64x256 dimensional embeddings. The architecture of such a model must efficiently blend the image and prompt to generate a mask, leading to the creation of three components: an Image Encoder, a Prompt Encoder, and a Mask Decoder. Since multiple prompts can be applied to the same image in an interactive scenario, both the Prompt Encoder and the Mask Decoder need to be swift and lightweight, while the Image Encoder can be more intensive. The Prompt Encoder outputs 256-dimensional embeddings for points, bounding boxes, and text prompts. The Mask Decoder is built with self-attention, cross-attention blocks, some MLP layers, and transpose convolution layers.

Unlike language models that have access to vast amounts of text data online, visual models face a scarcity of large-scale training data. To address this, Meta created a data engine, operating in three stages to create the SA-1B dataset. This dataset contains over 1.1 billion prompt-mask pairs, reflecting the back-and-forth between the model and annotators. Meta's SAM represents a pivotal leap in this direction.

In our approach, for the sake of simplicity, we used the sketch technique to specify the foreground area we intended to segment, employing points and bounding boxes. Given that most images in our study feature only one breast tumor, and that these tumor regions are typically contiguous, we limited ourselves to using a single prompt set per image. Following this, the SAM model's prompt encoder unit interprets the prompt, producing a fixed-length embedding that encapsulates the prompt's semantic context. This embedding is then merged with the output from the image encoder component to form a series of feature maps, which are then used by the mask decoder component to produce segmentation masks. We select the mask with the highest probability as our final segmentation mask. The prompts, ground truth mask contour and prediction mask contour of a sample image are shown in Figure 2.

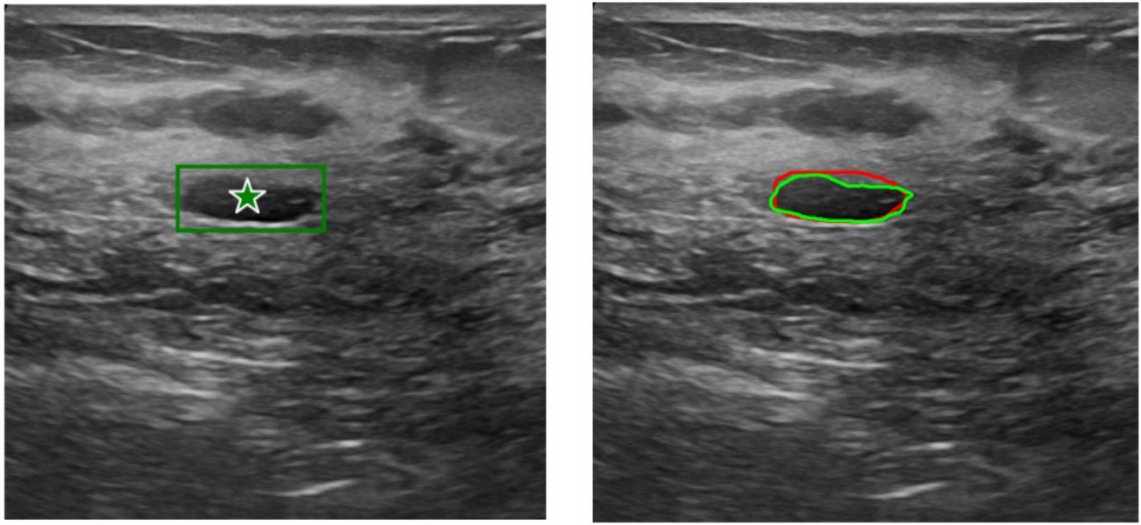

**Figure 2**: The left side of the image shows a sample image with prompts overlaying. A green bounding box is placed to indicate the intended segmentation area. Additionally, a green star is used to mark the foreground region. On the right side, the segmentation ground truth is displayed with a red contour representing the actual breast tumor. The predicted region is shown with a green contour.

2.3 Prompts Simulation

However, when it comes to evaluation on a large-scale dataset, hand-drawing prompts is practically infeasible. As a result, in order to mirror the process of generating prompts, we start by developing "ground prompts" based on the pre-existing segmentation masks and subsequently introduce a certain amount of randomness. The process unfolds as follows: We begin by accessing each image in the dataset along with its corresponding ground truth mask. Then, we randomly choose a point within the mask. The bounding box's dimensions are determined by the dimensions of the mask, with an additional 20 pixels on all sides. Following that, we randomly displace the bounding box up to 30 pixels in any direction, both horizontally and vertically. We also randomly scale the bounding box, allowing for up to a 10% increase. Finally, the bounding box and point are superimposed on the image to serve as prompts.

2.4 Evaluation Metrics

To assess the efficacy of our segmentation model, we utilized a variety of metrics, such as pixel accuracies, Intersection over Union (IOU), and the Dice Score. Pixel accuracies gauge the proportion of pixels that the model accurately categorizes. IOU quantifies the overlap between the predicted and actual masks by dividing the intersecting area of the masks by the combined area of the masks. The Dice Score offers another perspective on the overlap between the predicted and ground truth masks, computed by determining the harmonic mean of precision and recall scores. The definitions of these various metrics are displayed in Table 1.

Table 1. Evaluation metrics we used for this study and their definitions.

| Metric | Formula |
|---|---|
| Pixel Accuracy | Number of correctly predicted pixels / Total number of pixels |
| IOU (Jaccard) | Intersection over Union = Intersection / (Prediction + Ground Truth - Intersection) |
| Dice Score | 2 * Intersection / (Prediction + Ground Truth) |

## 3. Results

3.1 Comparison of Models

We analyzed and compared the performance of three pretained model varaints: ViT_h, ViT_l, and ViT_b. The ViT_h model demonstrated commendable performance with a mean pixel accuracy of 0.9609, a mean Dice score of 0.8291, and a mean IoU score of 0.7239. The ViT_l model displayed slightly superior performance compared to ViT_h in all categories. It exhibited a mean pixel accuracy of 0.9661, a mean Dice score of 0.8392, and a mean IoU score of 0.7361, suggesting its greater proficiency in segmenting images accurately. The ViT_b model, while still effective, showed the lowest performance among the three models in terms of mean scores, with a mean pixel accuracy of 0.9524, a mean Dice score of 0.8133, and a mean IoU score of 0.7020. All models exhibited proficient segmentation capabilities, but ViT_l standed out as the most effective model in terms of mean pixel accuracy, Dice score, and IoU score. Figure 3 visualizes the performance comparison of the three model variants.

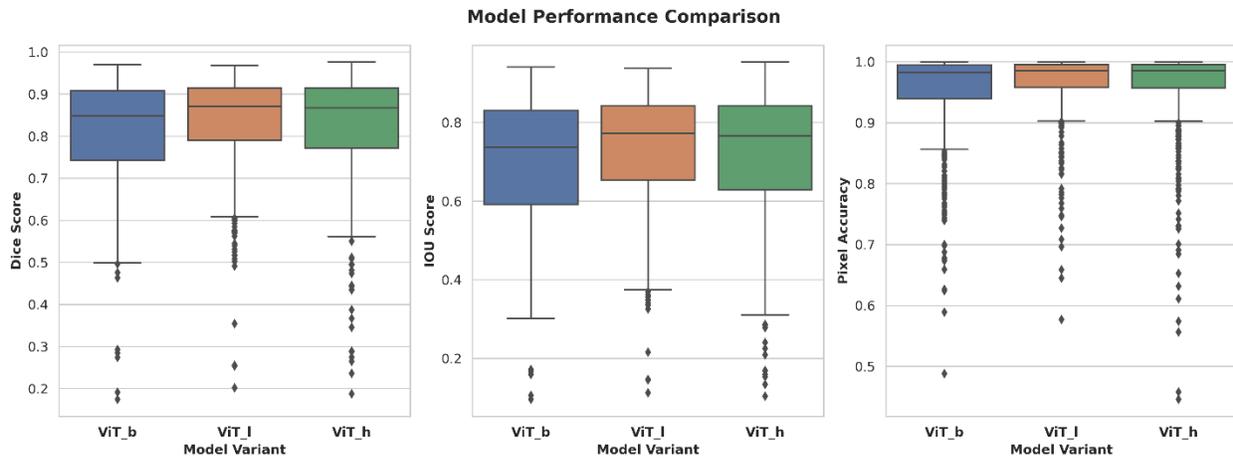

**Figure 3**: Performance Comparison of Different Model Variants: ViT_b, ViT_l, and ViT_h. The figure comprises three box plots, each representing the distribution of Dice Score, IOU Score, and Pixel Accuracy metrics for the three model variants respectively. Within each plot, the individual model's scores are displayed, allowing direct comparison. The box plot captures the interquartile range of scores with the median value represented by the line within the box. The model ViT_l has the best overall performance.

The discrepancy in performance between the ViT_h and ViT_l models, despite ViT_h having more parameters, can primarily be attributed to the intricacies of the optimization process during training. Models that are more complex, reflected in the larger number of parameters like those in ViT_h, are often linked with a more complicated loss landscape. This complexity can make it harder for optimization algorithms to effectively locate the global minimum, potentially impacting overall performance. Furthermore, it's important to note that these models were pre-trained on natural images. However, our study's focus was on breast ultrasound images, which have distinct characteristics compared to natural images. By maintaining the model weights from the pre-training phase, we essentially applied the knowledge from natural images directly to our medical imaging task. This process, while not exactly transfer learning, still involves a shift in data domains. The greater complexity of ViT_h, with its higher number of parameters, might have made it more difficult for the model to adapt its learned knowledge to the new domain of breast ultrasound images. On the other hand, ViT_l, despite having fewer parameters, may have been better able to apply its pre-trained knowledge to this new domain, resulting in better performance. For the rest of the experiments we used the best model ViT_L as the default model.

3.2 Prompt Interaction vs. No Prompt

In this experiment, we investigated the impact of prompt interaction on the segmentation performance of our model. Without the use of prompts, the model lacks crucial information about the user's intent, which results in difficulty in determining the precise areas to segment. When the model was trained without any prompts, we observed a significant drop in performance metrics. The mean pixel accuracy was only 0.1009, indicating that the model struggled to accurately identify the tumor regions. Similarly, the mean dice score was 0.0914, implying poor agreement between the predicted and ground truth masks. The mean IoU score further confirmed the limited overlap between the predicted and ground truth masks, with a score of 0.0540. In contrast, by incorporating prompt interaction into the training process, we enabled the model to understand the user's intent and better focus on the relevant areas for segmentation. Using the ViT_L model with prompt interaction, we observed substantial improvements in performance. The mean pixel accuracy significantly increased to 0.9661, demonstrating a significant enhancement in accurately identifying the tumor regions. The mean dice score improved to 0.8392, indicating a considerable improvement in the alignment between the predicted and ground truth masks. Furthermore, the mean IoU score reached 0.7361, highlighting a substantial increase in the agreement between the predicted and ground truth masks. Figure 4 visualizes the performance of ViT_l model with and without the prompt instructions.

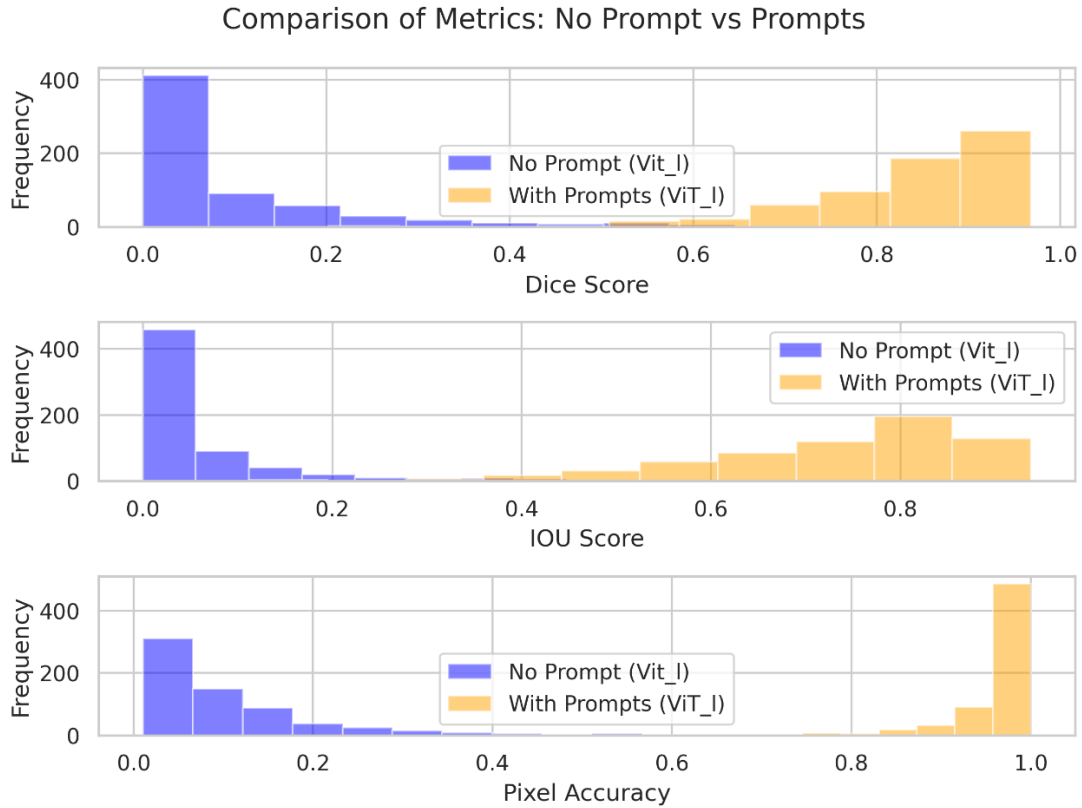

**Figure 4**: Comparison of Metrics: No Prompt vs Prompts. This figure compares the performance of the ViT_l model with and without prompt instructions. Histograms represent the distribution of three metrics: Dice Score, IOU Score, and Pixel Accuracy. Blue histograms depict results without prompts (Vit_l model), while orange histograms represent results with prompts (ViT_l model). The histograms visually demonstrate the impact of prompts on model performance. Prompts significantly improved all metrics, resulting in right-shifted and higher bars in the orange histograms compared to the blue histograms. Prompt instructions greatly enhanced the ViT_l model's segmentation accuracy and quality.

These findings emphasize the necessity of prompts for effective segmentation. Without prompts, the model lacks the contextual information required to understand the user's intent and identify the precise regions of interest. Prompt interaction enables the model to align its segmentation with the user's expectations, resulting in improved accuracy and better overall performance.

3.3 Malignant vs. Benign

In this section, we evaluated the model's segmentation performance in distinguishing between malignant and benign breast tumors, providing insights into its differential performance across these two categories.

For benign cases, the model exhibited exceptional segmentation accuracy, achieving a mean pixel accuracy of 0.9795. The mean dice score, quantifying the overlap between the predicted and ground truth masks, was 0.8567, indicating a substantial agreement between the model's segmentations and the actual boundaries of benign tumor regions. Additionally, the mean IoU score reached 0.7628, further confirming a significant overlap between the predicted and ground truth masks for benign cases.

In comparison, the model demonstrated commendable segmentation performance for malignant cases as well. It attained a mean pixel accuracy of 0.9382, reflecting accurate delineation of malignant tumor boundaries. The mean dice score of 0.8029 indicated a notable overlap between the predicted and ground truth masks, further corroborating the model's ability to capture the malignant tumor regions accurately. The mean IoU score of 0.6805 indicated a reasonable agreement between the predicted and ground truth masks for malignant cases.

These results highlight the model's efficacy in distinguishing and segmenting malignant and benign breast tumors. The higher pixel accuracy, dice score, and IoU score achieved for benign cases suggest a superior ability to accurately delineate the boundaries of benign tumors. Figure 5 shows the difference of performance between segmenting the benign and malignant breast tumors.

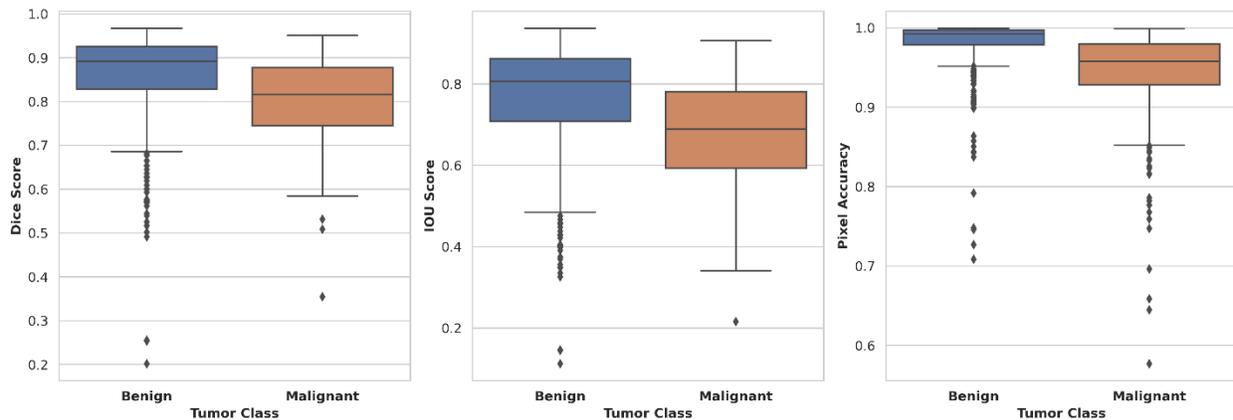

Figure 5: Benign vs Malignant. This figure presents the performance difference between segmenting benign and malignant breast tumors. Each subplot represents a different metric, including Dice Score, IOU Score, and Pixel Accuracy. In each subplot, boxplots illustrate the distribution of the respective metric for both benign and malignant tumor classes. The x-axis represents the tumor class, with "Benign" and "Malignant" labels. The y-axis corresponds to the metric values, indicating the segmentation performance.

The disparity in segmentation performance between benign and malignant breast tumors can be attributed to the following factors. Firstly, benign tumors often possess well-defined and distinct boundaries, which facilitates their accurate segmentation. These tumors tend to exhibit a more regular and organized structure, allowing the model to effectively capture and delineate their boundaries. In contrast, malignant breast tumors are characterized by greater heterogeneity, irregular shapes, and boundaries. Their aggressive nature and invasive growth patterns pose challenges for accurate segmentation. The increased complexity and variability associated with malignant tumors make it more difficult to capture their full extent, resulting in slightly lower segmentation performance. The intricate nature of malignant tumors, combined with their aggressive behavior, adds complexities to the segmentation process, which affects the model's ability to accurately identify and delineate their boundaries. Thus, the variations in tumor characteristics and complexities between benign and malignant breast tumors contribute to the observed differences in segmentation performance.

3.4 Impacts of Breast Tumor Characteristics

In this section, we explore various characteristics of breast tumors and their potential impacts on tumor segmentation. Understanding these characteristics is essential for analyzing segmentation performance and developing robust algorithms. The following tumor characteristics are considered:

- Tumor Size: Tumor size refers to the physical dimensions of the tumor within the breast. It represents the extent of the tumor in terms of the number of pixels or its spatial dimensions. Larger tumors occupy a greater area within the breast, while smaller tumors occupy a relatively smaller area.
- Tumor Contrast: Tumor contrast relates to the visual distinction between the tumor region and the surrounding background tissues in medical images. It encompasses variations in intensity, color, or texture that differentiate the tumor from its surroundings. Tumors with higher contrast exhibit more pronounced visual differences, while tumors with lower contrast may blend with surrounding tissues, making segmentation more challenging.
- Tumor Aspect Ratio: The aspect ratio characterizes the shape of the tumor by representing the ratio of its length to its width or height. It provides information about the elongation or distortion of the tumor shape.
- Tumor Complexity: Tumor complexity relates to the intricacy and irregularity of the tumor boundaries and internal structures. It encompasses factors such as irregular shapes, fragmented patterns, or heterogeneous texture. Complex tumors exhibit more intricate characteristics, while less complex tumors have more regular and easily defined boundaries.

To assess the complexity of a tumor, we employed a Fourier analysis-based method. The method involved analyzing the contour of the tumor, which represents its boundary. By applying the Fast Fourier Transform (FFT) to the contour, we obtained Fourier descriptors that captured the frequency components of the tumor's shape. To quantify the tumor's complexity, we computed the ratio of the magnitude of the first Fourier descriptor to the sum of the magnitudes of the remaining descriptors. This ratio provided a measure of the tumor's intricacy and irregularity. Higher complexity values indicated more complex and irregular tumor shapes, while lower values suggested relatively simpler shapes.

We conducted a correlation analysis using Kendall's Tau to examine the relationships between the segmentation performance and various tumor characteristics. The results revealed interesting insights into the associations between these factors. We observed a negative correlation between the tumor's complexity and segmentation performance. As expected, these complexities pose challenges for accurately capturing tumor boundaries and differentiating them from surrounding tissues, resulting in decreased segmentation performance. The tumor size has a minor positive correlation with the segmentation performance. Larger tumors often exhibit more distinct boundaries, providing clearer visual cues for segmentation algorithms. Additionally, the presence of a larger tumor can offer more context and information, aiding the segmentation process and contributing to improved performance. The tumor contrast has the highest correlation among all characteristics. Tumors with higher contrast exhibit more pronounced visual differences, which make it easier for segmentation algorithms to accurately delineate their boundaries. The clearer and more distinct boundaries provided by higher contrast tumors enable algorithms to better differentiate and separate the tumor regions, leading to a higher correlation with segmentation performance. And aspect ratio barely impacts the performance. Figure 6 is the correlation matrix of the Kendall's Tau correlation analysis. Figure 7 uses scatterplot to visualize the correlation between the segmentation performance and various tumor characteristics.

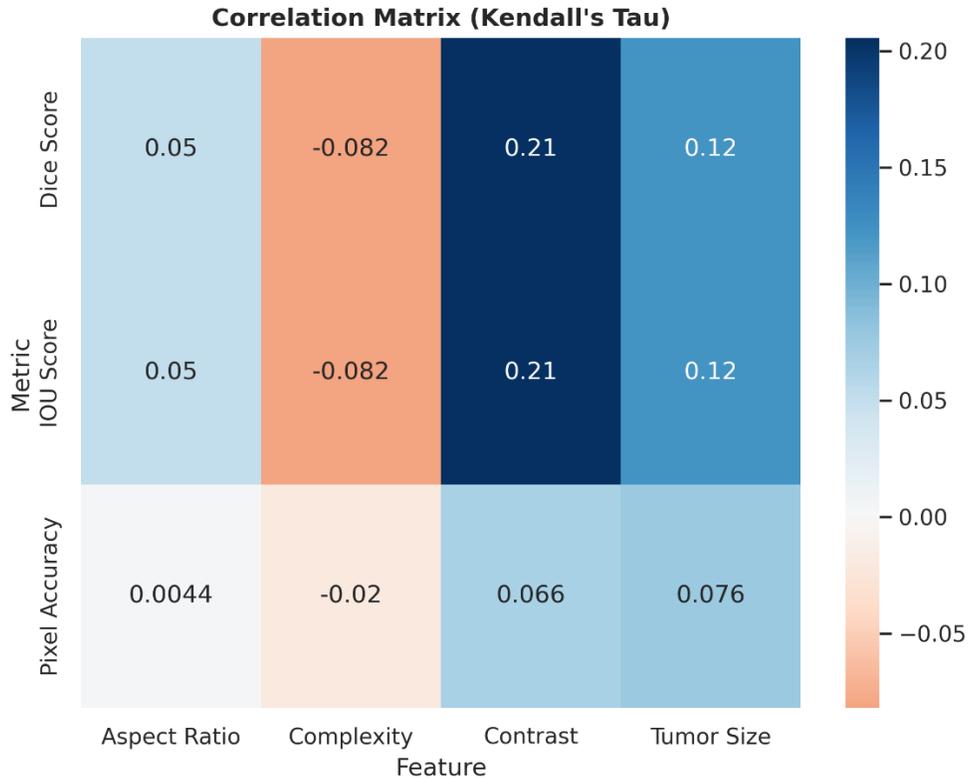

Figure 6. Kendall's Tau correlation matrix. We Can observer that better contrast larger tumor size can lead to better segmentation result while it is usually harder to segment out tumors with complex boundaries.

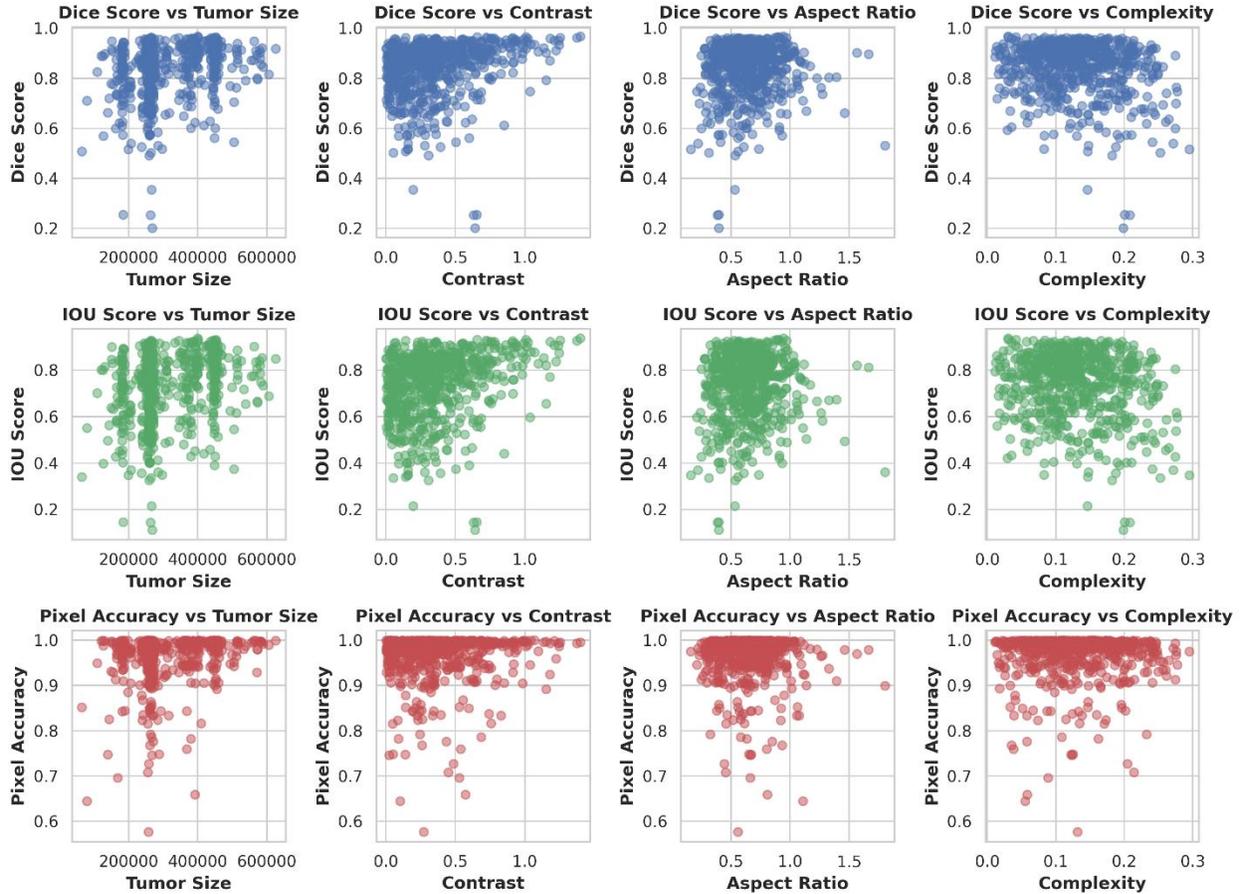

**Figure 7**. This figure showcases scatter plots examining the relationships between segmentation performance metrics (Dice Score, IOU Score, and Pixel Accuracy) and various tumor characteristics (Tumor Size, Contrast, Aspect Ratio, and Complexity). Each scatter plot represents a specific metric against a specific characteristic. The plots demonstrate the distribution of data points and provide visual insights into the correlations between the metrics and characteristics.

## 4. Discussion and Conclusion

In this study, the successful application of SAM for breast tumor segmentation has demonstrated its effectiveness in the medical imaging domain. SAM, originally designed as a powerful visual model for natural image segmentation, has showcased its potential to be a valuable tool in medical image analysis. The remarkable performance achieved in this study highlights the adaptability and versatility of SAM in handling complex medical images.

However, it is important to acknowledge the existing room for improvement in our research. While the obtained results were highly promising, it is crucial to validate the robustness of the model on multiple datasets. By testing SAM on different breast tumor ultrasound datasets, we can assess its generalizability and verify its performance across various patient populations.

Moreover, our evaluation focused on utilizing a pre-trained model without fine-tuning. Fine-tuning the model using a larger and more diverse dataset has the potential to further optimize its segmentation capabilities. The collection of additional data will enable us to refine the model's parameters and adapt it more specifically to the characteristics of breast tumor ultrasound images. This iterative process of fine-tuning holds the promise of enhancing SAM's performance and achieving even better segmentation results.

Furthermore, while our current analysis was conducted on 2D images, it is essential to consider the prevalence of 3D medical images in clinical practice. Segmenting 3D medical images slice by slice, as commonly practiced, may result in information loss and reduced accuracy due to the lack of context between adjacent slices. To overcome this limitation, future work will focus on extending SAM to 3D image segmentation, enabling a more comprehensive analysis of volumetric medical data. This expansion will unlock opportunities for more accurate tumor characterization and treatment planning.

Beyond breast tumor ultrasound images, the application of SAM can be extended to other medical image modalities, such as MRI, CT, and histopathology slides. By adapting SAM to different imaging modalities, it is possible to develop a unified framework for tumor segmentation and analysis across multiple medical domains. This interdisciplinary approach holds immense potential for advancing the field of medical image analysis and facilitating improved clinical decision-making.

In conclusion, we compared the performance of three pre-trained model variants: ViT_h, ViT_l, and ViT_b. Among these models, ViT_l exhibited the highest performance, surpassing ViT_h in all categories, including mean pixel accuracy, mean Dice score, and mean IoU score. Furthermore, when examining the segmentation performance in distinguishing between malignant and benign breast tumors, the model displayed exceptional accuracy for both tumor types. Benign tumors exhibited higher pixel accuracy, Dice score, and IoU score, indicating better segmentation results compared to malignant tumors. This observation can be attributed to the distinct characteristics of benign tumors, such as well-defined boundaries and regular structures, which facilitate accurate segmentation. Malignant tumors, on the other hand, with their greater complexity and irregularity, pose challenges for precise boundary delineation. Regarding the impacts of breast tumor characteristics, we analyzed tumor size, contrast, aspect ratio, and complexity. The results revealed that tumor contrast had the highest correlation with segmentation performance, as tumors with higher contrast displayed more pronounced visual differences, making their boundaries easier to delineate. Tumor size demonstrated a minor positive correlation, suggesting that larger tumors with clearer boundaries can contribute to improved segmentation performance. Complexity showed a negative correlation, indicating that tumors with more intricate boundaries and irregular shapes presented challenges for accurate segmentation. The aspect ratio had a negligible impact on segmentation performance.

While our study demonstrates the remarkable performance of SAM in breast tumor segmentation, there are opportunities for further research and improvement. By addressing the identified limitations, including dataset diversity, fine-tuning, and expanding SAM to 3D and other imaging modalities, we aim to advance the capabilities of SAM and contribute to the broader field of medical image segmentation. Ultimately, our efforts strive to improve the accuracy, efficiency, and reliability of tumor analysis, leading to enhanced patient care and outcomes.